# *In-situ Nanoscale Characterization of Composition and Structure during Formation of Ultrathin Nickel Silicide*

Tuan T. Tran [a], Christian Lavoie [b], Zhen Zhang [c], and Daniel Primetzhofer [a]

[a] *Department of Physics and Astronomy, Ångström Laboratory, Uppsala University, Box 516, SE-751 20 Uppsala, Sweden*

[b] *IBM Thomas J. Watson Research Center, Yorktown Heights, New York 10598, USA*

[c] *Solid State Electronics, Department of Electrical Engineering, The Ångström Laboratory, Uppsala University, SE-75121 Uppsala, Sweden*

Corresponding author: Tuan Tran (tuan.tran@physics.uu.se)

## Abstract

We characterize composition and structure of ultrathin nickel silicide during formation from 3 nm Ni films on Si(100) using *in-situ* high-resolution ion scattering and high-resolution transmission electron microscopy. We show the transition to occur in discrete steps, in which an intermediate phase is observed within a narrow range of temperature from 230 ºC to 290 ºC. The film composition of this intermediate phase is found to be 50% Ni:50% Si, without evidence for long-range structure, indicating the film to be a homogeneous monosilicide NiSi phase. The final phase is resemblant of the cubic disilicide $NiSi_2$, but with slightly off-stoichiometric composition of 38% Ni and 62% Si. Along the [100] axis, the lattices of the film and the substrate are found in perfect alignment. Due to the epitaxial growth of the silicide, a contraction of the *c* lattice constant of the film by 0.7 – 1% is detected.

## Keywords:

*Ultrathin, nickel silicide, epitaxial silicide, in-situ characterization, high-resolution ion scattering.*

# Introduction

Nickel (Ni) silicide is an important local contact material for silicon (Si) transistors due to its low resistivity, immunity to the fine-line effect and low Si consumption [1]. At increasing annealing temperature, the normal phase transition sequence and the formation temperature of nickel silicides formed from Ni films with thickness > 4 nm is δ-$Ni_2Si$ (~250 °C) → NiSi (~350 °C) → $NiSi_2$ (~800 °C) [1,2]. The intermediate phase, nickel monosilicide (NiSi), is usually desirable for high-performance devices. Several studies have, however, suggested that the transition of the Ni-Si system is more complex than the thermodynamic sequence because a metastable silicide, namely the θ phase, plays an important role in the transition. The θ phase has a hexagonal structure with varied Si composition (33 − 40 at. % Si). Thermodynamically, it is stable at temperature > 825 °C. However, due to texture inheritance from the δ-$Ni_2Si$ phase it is metastable at around 300 °C. The presence of the θ phase then facilitates the transition to the NiSi phase because this process requires little movement of the atoms from their equilibrium positions [3,4].

The thickness of the initial deposited Ni films also vastly modifies the transition sequence of the silicides. Thinner films are more prone to agglomeration due to capillary effects [5]. Reports have shown that the NiSi phase, formed from the Ni films with initial thickness of 4 − 15 nm, becomes agglomerated at 500 − 600 °C, leading to sharp increase in resistivity of the films [1,5]. However, for the Ni film thickness < 4 nm the silicide film becomes very stable and strongly against agglomeration, up to a temperature of at least 800 °C [6-8]. This phenomenon was first reported by Tung *et al.* [9], in which the silicide layer was shown to be perfectly epitaxial on either Si(100) or Si(111) with low interfacial defect densities. The epitaxial films form at relatively low temperature (350 − 450 °C), and therefore, strongly stabilize the silicide layer against agglomeration.

The understanding on the phase transition of this ultrathin silicide ($t < 4$ nm) is not well established or under debate. For example, there is disagreement among reports on the composition and the phase of the epitaxial layer. Using Auger electron spectroscopy, Tung *et al.* showed that the composition of the epitaxial phase is nickel disilicide, $NiSi_2$, based on the peak-to-peak height ratio between the Ni *LMM* and the Si *LVV* peaks. However, a study by Luo *et al.* using high-resolution transmission electron microscopy (HR-TEM) showed that the thickness ratio between the silicide and the initial Ni thickness was just 2.6 − 2.7, considerably below the conversion ratio of the bulk $NiSi_2$ phase of 3.62 [5,10]. Another study, using a combination of energy dispersive X-ray spectroscopy, TEM, electron backscatter diffraction and pole-figure measurement, showed that no single epitaxial phase can explain all the data. The epitaxy-like layer is rather a mixed phase of small domains (a few nanometer) that share alignment of the atomic planes [6].

Future applications, such as the extremely-scaled field-effect transistors or the nanowire devices [11-13], will rely on electrical contacts of a few nanometers in thickness [14]. Therefore, a thorough understanding of the entire formation process of such ultrathin silicides is essential. In this study, we employed high-resolution medium-energy ion scattering (MEIS) with a large-are position-sensitive detector to study *in-situ* the phase transition of the silicide. This non-destructive technique can provide simultaneously depth-resolved composition and real-space crystallography of the film throughout the sequential annealing [15]. Complementarily, we used high-resolution transmission electron microscopy to characterize the crystal structure, defects and lattice strains of the resulting silicide film.

## Experiments

Ni films were deposited on p-type $1 - 10$ $\Omega$ cm Si(100) substrates using magnetron sputtering. Prior to the film deposition, the Si surface was cleaned using 0.5% hydrofluoric acid and then immediately loaded into the sputtering system. A nominal Ni thickness of 3 nm was set for the deposition by timing. Accurate areal thicknesses of the films were, however, measured using Rutherford backscattering spectrometry (RBS) with a 2 MeV helium (He) beam. The RBS showed the areal density of the film is $\sim 25 \times 10^{15}$ at/cm$^2$. Assuming the deposited films having bulk density of 8.9 g/cm$^3$, the physical thickness, as calculated from the areal density, is 2.7 nm. The phase transition study was done *in-situ* using a 3D-MEIS system at Uppsala University [16,17]. A noticeable feature of this system is a multi-channel-plate delay-line detector (RoentDek DLD120) coupled with a time-of-flight mechanism. This system allows us to acquire the energy and the position on the detector of every single detected ion. Hence, depth-resolved information on both composition and crystallography can be achieved simultaneously [15]. Although using energetic ions, the technique can be considered non-destructive for the following reason. The dose required for the measurements is in the order of $10^7$ at/cm$^2$, creating a total number of vacancies in the order of $10^8$ at/cm$^2$ over a depth of 10 nm, according to a SRIM simulation [18]. Considering the areal density of the $\sim 10$ nm film in the order of $10^{16}$ at/cm$^2$, the total induced displacement is at most ten billionth of the film. Since, this damage is minimal, *in-situ* studies using a single sample is possible throughout the process. Other features of the system include a 6-axis goniometer (3 translational and 3 rotational axes) and a heating filament below the samples. Temperature monitoring was done using a k-type thermocouple attached directly to the sample surface. This thermocouple was calibrated using the melting points of indium ($t_m = 156.6\,°C$) and lead ($t_m = 327.5\,°C$) as references.

MEIS measurements were done with beams of He ions at primary energies of $50 - 200$ keV. The low energy measurement provides better depth resolution for the determination of the composition. In contrast, the higher energy beam enables better structural characterization. Several different scattering geometries were employed. Random geometries are used for composition analysis; in such sample normal and detector axis are oriented to avoid any low-index crystal axes and planes. This geometry thus avoids uncertainties related to channelling and blocking, which is necessary to employ the SIMNRA simulation program [19]. Real-space crystallographic studies were effectively conducted with close alignment of the detector to a major axis, such as the [100] and the [101] directions. Blocking patterns of these axes, together with the energy discrimination of detected particles reveal the crystal structure of the films and the substrates. Blocking patterns also contain depth resolved information on the lattice strain usually associated with ultrathin epitaxial films. The size of the beam on the sample is about $0.5 \times 5$ mm. The position of the beam was kept fixed in all measurements. During annealing, the sample temperature was increased at a rate of $\sim 20\,°C/\text{min}$ in average, kept on hold for 1 min at the desired temperature and then cooled down close to room temperature for the ion scattering measurement. After the ion scattering experiments, the resulting samples annealed at $540\,°C$ were further characterized with HR-TEM. Cross-sectional lamellae of the samples were prepared with a focused ion beam system. Whereas, the HR-TEM micrographs were acquired with a FEI Titan Themis 200 system at the electron energy of 200 keV. Analysis of the TEM micrographs was conducted with the Gatan's Digital Micrograph software (GMS3). This software allows us to implement forward and backward Fourier transform of the HR-TEM micrograph to assess the crystal defects and lattice strains.

## Results

Fig. 1 (a-c) show MEIS spectra acquired in backscattering geometry for the sample annealed sequentially to increasing temperature from as-deposited to $540\,°C$. Fig. 1 (d-f) are the [100] blocking patterns that are purposely selected to show the onset of the crystallization. It is worth noting that the blocking patterns shown consist of ions having a final energy of $34 - 42$ keV only, to limit the signal to scattering from Ni atoms. Hence, they yield information on the structure of the silicide film exclusively, not obscured by the substrate signal. Fig. 1a shows that the first subtle changes in the film composition occurred at 230 °C undergoing no further change up to 260 °C. Note that for random direction, in first order, both the height of the Si-substrate signal and the integral area of the Ni-signal (proportional to the amount of Ni-atoms) are expected to be constant and can be used for spectrum normalisation. Specifically, at 230 °C and 260 °C, the distribution of the Ni-signal is broadened on the expense of a reduced intensity indicating the first reaction between the film and the substrate. The blocking pattern for the 260 °C measurement (Fig. 1d) does not show any distinct pattern, supposedly the Ni film remained disordered (amorphous or polycrystalline) at this stage. The minimum visible at the lower edge is due to ions channelling in the first MCP of the detector stack, i.e. an artefact not associated with the sample.

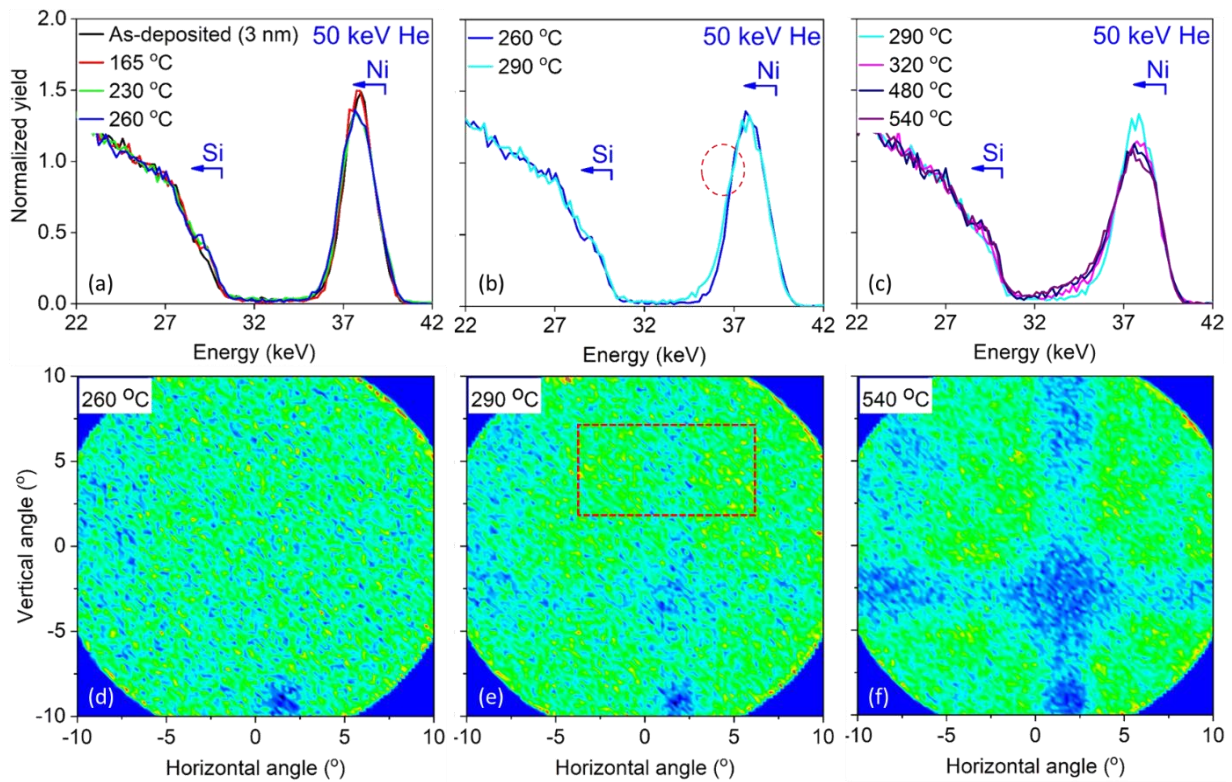

Fig. 1: Backscattering spectra (a-c) and the corresponding blocking patterns (d-f) selected to show the onset of the crystallization. All spectra and patterns were recorded with a 50 keV He ion beam. Backscattering spectra and blocking patterns were taken in random geometry and around [100] axis, respectively.

A pattern is first observed only at a slightly higher temperature of 290 °C (Fig. 1e). As compared to the 260 °C spectrum, the backscattering spectrum for the 290 °C in Fig. 1b is almost similar, except for a small shoulder (red-dash circle) and broader energy distribution at the low-energy edge of the Ni peak. This observation suggests that while inter-diffusion from Ni and Si can be observed from 230 ℃, a clear crystallization of the silicide film starts to occur at the interface at the annealing temperature of 290 ℃. Considering the total thickness of the Ni film of ∼3 nm, the epitaxial interfacial layer is estimated to be only a few monolayers. Still, those few monolayers are sufficient to block part of the intensity of the scattered beam and form the observed pattern. In supplementary section, we provide further analysis of the Fig. 1(e) blocking pattern. The blocking profiles within the red-dash box are compared for the first and the second half of the silicide film. The stronger blocking effect of the half film closer to the interface further supports that the crystallization initiates at the interface. Upon further annealing to a temperature of 320 ℃, a significant change in the composition of the film is observed (Fig. 1c). At this stage, the whole silicide film appears to have transformed and is subsequently only changing slightly as the temperature increases up to 540 ℃. The blocking pattern becomes much more apparent (Fig. 1e), suggesting an improved crystallization of the silicide films. The almost complete absence of scattered particles in the [100] direction, i.e. in the minimum located at 2° in both vertical and horizontal direction on our detector, is in fact an indication of the crystal structure being expressed throughout the whole film thickness.

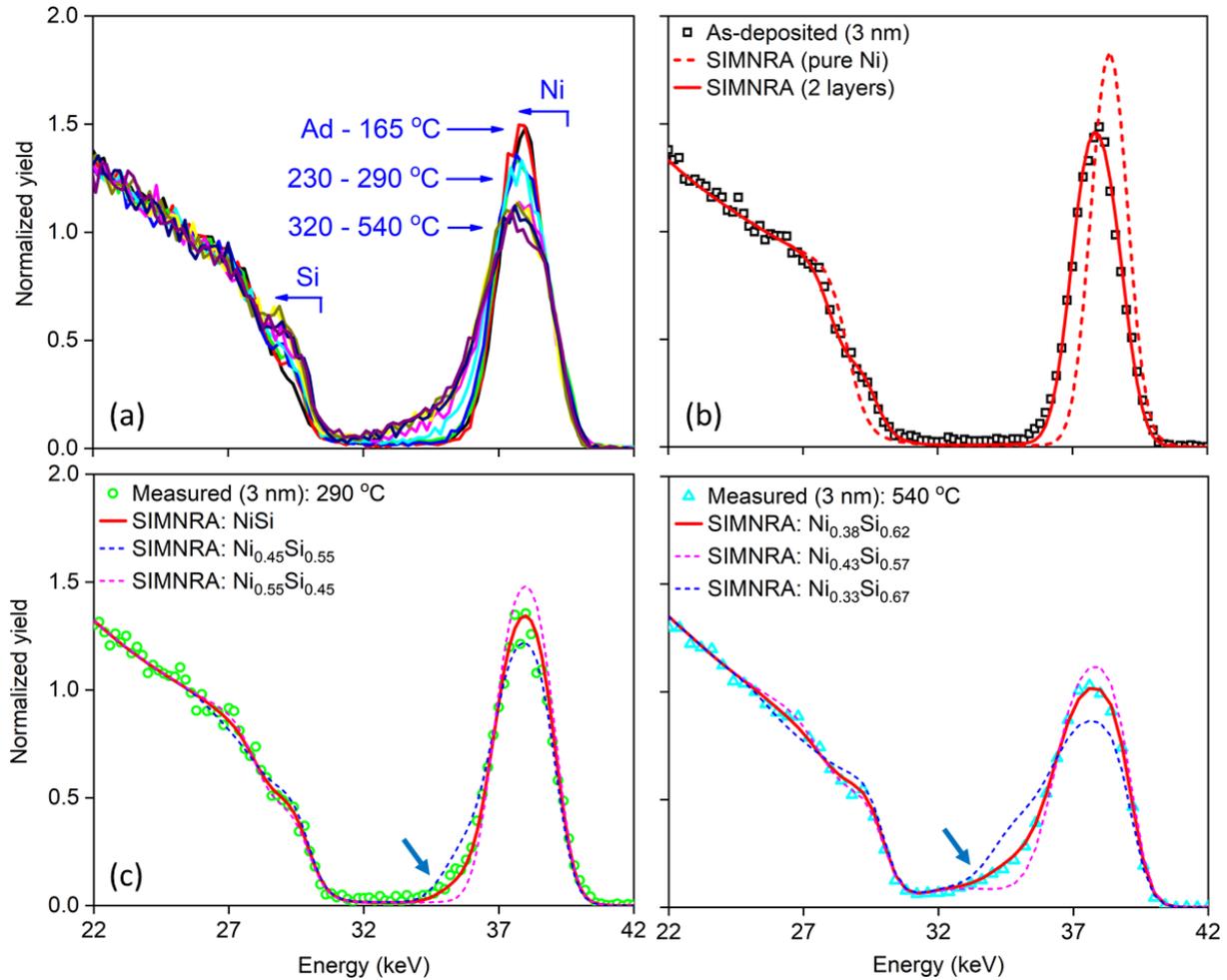

Fig. 2: Combination of the backscattering spectra in one graph (a). Simulations using SIMNRA are shown for the as-deposited, 290 °C and 540 °C spectra (b-d).

A combination of all backscattering spectra is shown in Fig. 2a. The yield of the spectra was normalized to the yield of the unaffected signal of the Si substrate. Regarding the Ni peaks, one can see three distinct stages of the transition: as-deposited - 165 °C, 230 − 290 °C and 320 − 540 °C. In other words, the sample did not transform abruptly from the as-deposited to the crystalline silicide film, but instead transforms in at least two discrete steps, in a similar manner than observed for thicker films. The film composition can be determined by simulating the backscattering spectra using SIMNRA as shown in Fig. 2b-d [19]. For reliable simulation, several constraints are applied. The number of the Ni atoms is determined using the well-calibrated RBS method and then fixed among the three spectra, as both evaporation and deep diffusion is severely unlikely. The energy resolution of the detector is also kept constant. Finally, the stopping cross sections of Ni, Si and nickel silicide, providing accurate depth scales were taken from a recent study using thicker films, which are more suitable to determine the accurate stopping values of the individual materials [20].

Fig. 2b shows the spectrum of the as-deposited sample. Although the sputtering was done at room temperature, an intermixing layer of Ni and Si can be observed at the interface of the Ni film and the substrate. As compared to the simulation using a pure Ni layer (dash-red curve), the one using a top $N_2Si$ layer and another layer of lower Ni concentration (red solid curve) gives a perfect fit to the measurement. Room temperature reaction of the deposited metals and silicon has been also reported both for sputtering and e-beam evaporation techniques [21,22]. For the intermediate phase (Fig. 2c), the simulation resulting in the best fit shows that the composition of the film is $50:50$, i.e. the monosilicide phase (NiSi) with the areal density of $47 \times 10^{15}$ at/cm$^2$. Assuming the NiSi film has bulk density of $5.93$ g/cm$^3$ [23], the calculated thickness is ~5.7 nm. A small percentage of Ni (in total of ~6 at.%) diffused into the substrate with a depth of ~6 nm. Also this diffusion is accounted for in the simulations, shown by the tail at around 35 keV. Included in the figure are two other simulated spectra with $\pm 5$ at.% variability to provide an assessment of the accuracy. The intermediate NiSi phase was observed only within a short range of temperature ($230 - 290$ °C), which makes it easy to be omitted in current literature. According to Fig. 1(d) no pattern was observed for the 260 °C measurement. Therefore, the monosilicide phase is presumably not a single-crystalline or epitaxial film.

Finally, Fig. 2d shows the simulation of the best fit and the $\pm 5$ at.% variability of the final silicide phase. The simulation of the stoichiometric $NiSi_2$ phase (dash-blue curve) appears to have lower Ni peak and higher Si edge than the measurement. Whereas, the best fit (solid-red curve) gives 38% Ni and 62% Si, i.e. slightly increased Ni-concentrations. The areal density of the film is $57 \times 10^{15}$ at/cm$^2$. To calculate the film thickness in nanometers, the mass density of the film is required. Since the film structure and composition is closely resemblant of the $NiSi_2$ phase, we use the density of this phase of $4.86$ g/cm$^3$ for the conversion [24]. As a result, the film thickness, excluding the diffusion tail, is 7.2 nm. A total fraction of 17% Ni atoms are also found diffusing to a depth of ~18 nm below the film.

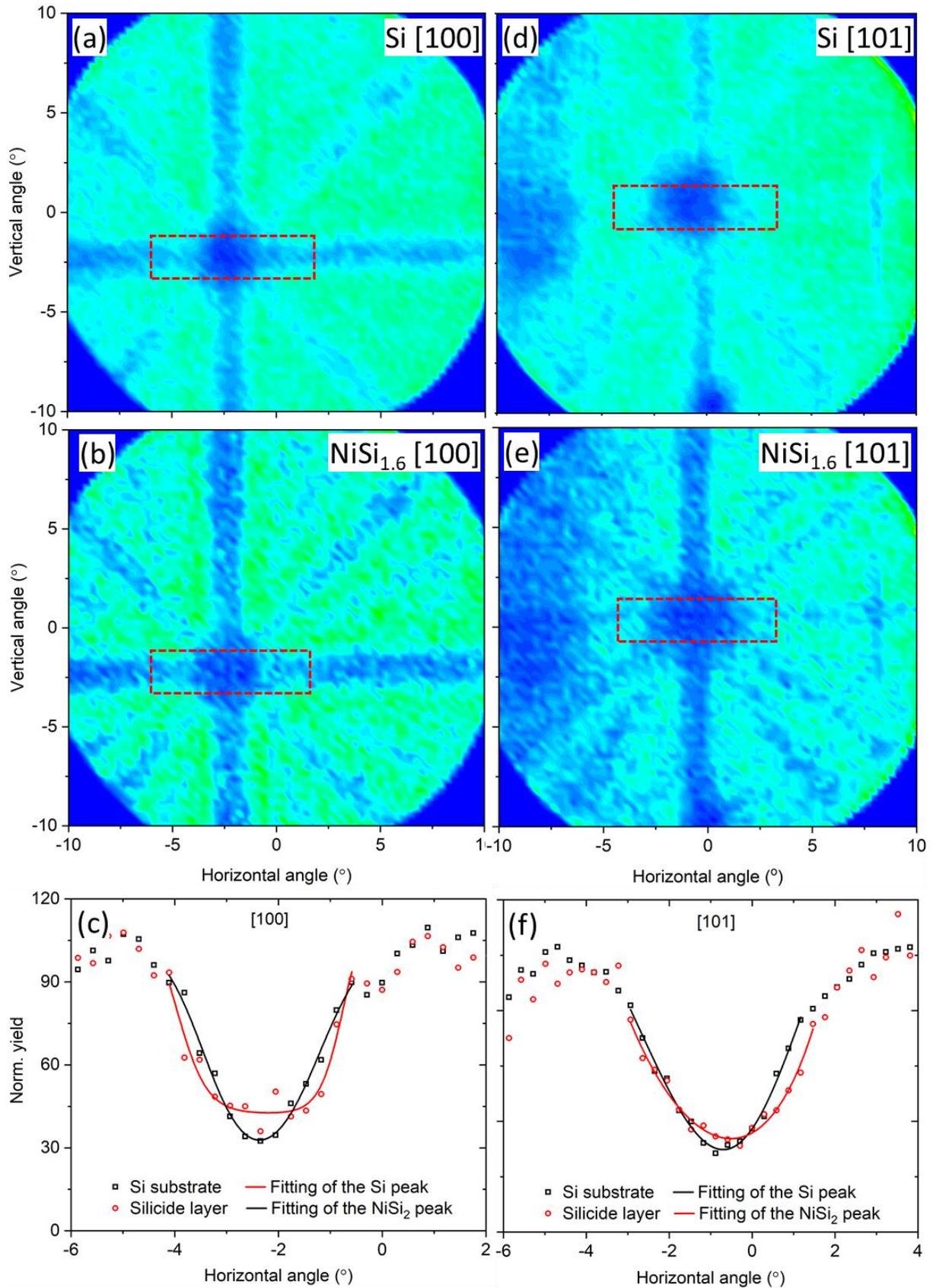

Fig. 3: Blocking patterns of the sample annealed at 540 °C around the [100] axis (a, b) and the [101] axis (d, e). The scattering yields of the substrate and the silicide, as integrated within the red-dash box, are presented for the [100] (c) and the [101] axis (f).

Further investigation on the crystal structure of the final silicide phase and its relation to the substrate is done by analysing the blocking patterns of both the film and the substrate in combination as shown in Fig. 3a, b. These two patterns were recorded simultaneously in one measurement using a 200 keV He beam and with the center of the detector oriented close to the [100] axis of the samples. The pattern (a) consist of detected He ions with the final energy of $80 - 110$ keV, i.e. scattered from the Si atoms of the substrate. Pattern (b) is obtained from ions scattered from the Ni atoms in the silicide films. The high degree of similarity for the two patterns, in terms of appearance, position of the main axis and the relative angles between the planes, suggest that the silicide film is single-crystalline, has highly similar structure and, moreover, is epitaxial with the Si substrate. This epitaxial growth of the silicide film, in fact, has been reported in many studies [5,6,25]. By integrating the scattered yield around the main [100] axis (dash-red region), more quantitative analysis is achieved as shown in Fig. 3c. The exact same locations of the silicon and the silicide peaks again confirm the commensuration between the two lattices with the [100] axes in perfect alignment. Fig. 3d is prepared by doing the same procedure as Fig. 3c, but with the detector center around the off-normal [101] axis of the sample. In this measurement geometry, a clear shift of the silicide peak with respect to the silicon peak is recognisable. From a fit to the data, the value of this shift results to $\sim 0.2°$ towards a larger exit angle. As the two patterns were acquired simultaneously in one measurement and one detector the only difference is the energy ranges of the ions by which the patterns are constructed, no uncertainty related to the measurement geometry is expected. Considering the crystalline silicide film having a cubic structure similar to silicon, a shift of $\sim 0.2°$ towards higher exit angle is equivalent to a contraction of the *c* axis of $\sim 0.7\%$.

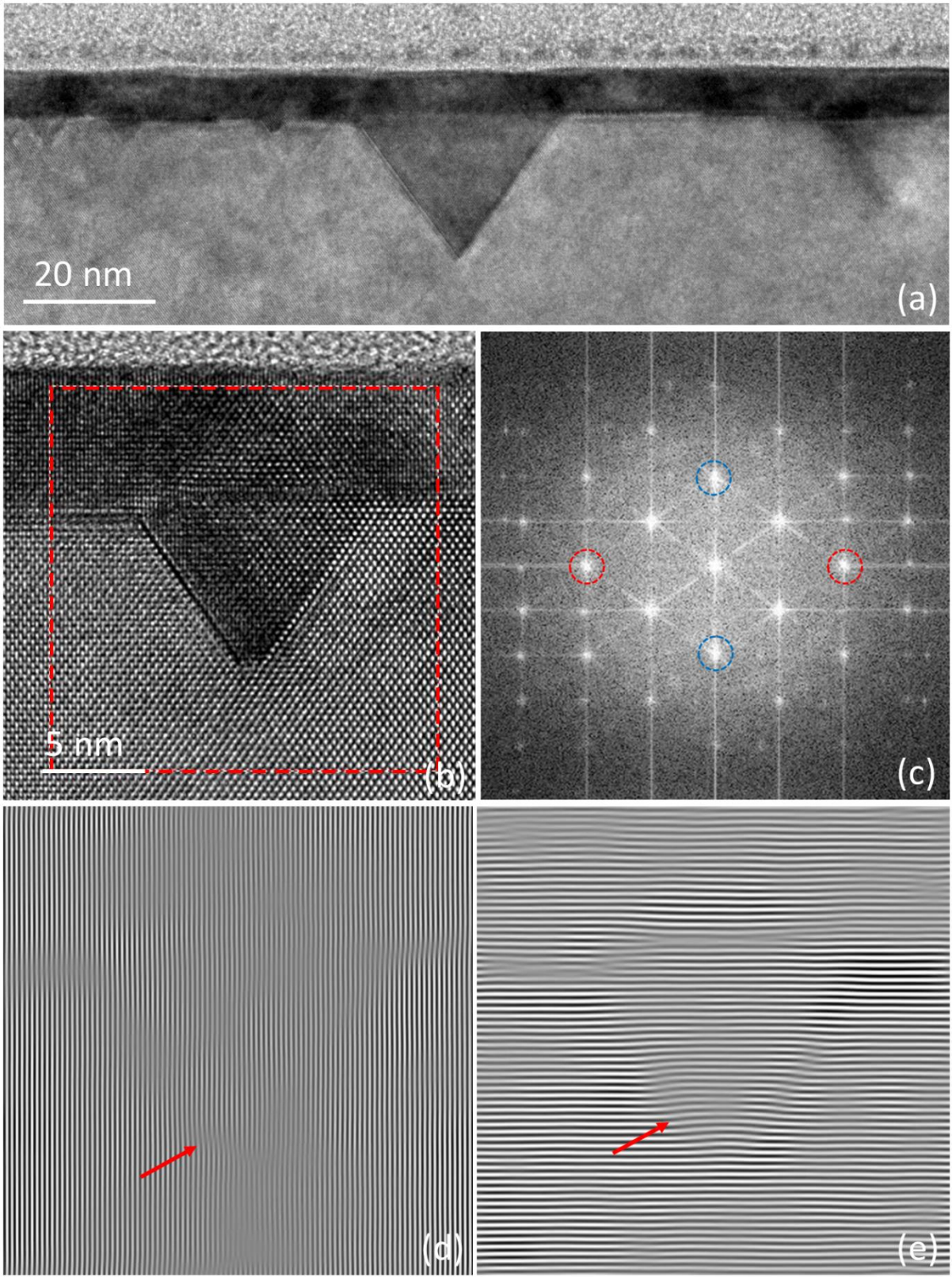

Fig. 4: Cross-sectional transmission electron micrographs of the Ni silicide sample annealed at 540 °C for 1 min (a). Fig. (b) is the high-resolution image of the pyramid region. Fig. (c) presents the Fourier transform within the red-dash square of Fig. (b). After filtering out other features, except the region in the red-dash circles, the filtered FT pattern is inversed

to generate the HR-TEM image of only the vertical lattice planes in the Fig. (d). Using the similar method, Fig. (e) was prepared to show the HR-TEM of the horizontal planes.

Complementary to the ion scattering study, cross-sectional transmission electron microscopy is employed to assess the crystallinity, defects and local strains of the films. The TEM lamella shown in Fig. 4 was taken from the sample annealed at 540 °C for 1 min. Fig. 4a shows that the silicide film is continuous, flat and has no agglomeration. However, below the films pyramids with different sizes are also observed. The existence of such pyramids is expected to account for the observation of the diffusion tails in the backscattering spectra in Fig. 2. Based on the combined ion scattering and the TEM results, one can describe the transformation process of the ultrathin silicide films as follows. The as-deposited film remains largely unchanged until 230 °C. At this temperature, it transforms to the non-epitaxial monosilicide phase and is stable only within a narrow range of temperature (230 − 290 °C). The early crystallization of the final silicide, presumably from a few continuous monolayers, occurs at the interface at 290 °C. This interfacial seed then triggers the complete crystallization of the whole film shortly after, at the temperature of 320 °C. At this temperature, a fraction of the Ni atom (∼ 6 at. %) starts to form the pyramids. At higher temperature, the pyramids grow in size as the ion scattering indicates ∼ 17 at. % of Ni diffusing below the film to a depth of up to ∼ 18 nm.

The HR-TEM image around the pyramid area shows three distinct regions: the film, the pyramid and the substrate. The thickness of the film, as measured from the image, is ∼6.0 − 7.0 nm, which is slightly smaller than 7.2 nm as measured from the ion scattering. However, since the mass density used for the ion scattering measurement is for the stoichiometric $NiSi_2$, this discrepancy in the thickness between the TEM and the ion scattering further supports the evidence that the epitaxial silicide phase is not fully stoichiometric, but composed of slightly elevated Ni concentration, such as 38 % Ni and 62 % Si. Also, both size and shape of the pyramids is in good agreement with the Ni-interdiffusion employed in the simulations

Further image processing is done using Fourier Filtering and Reconstruction as described elsewhere [26]. This procedure allows us to disentangle different lattice planes making up the high-resolution image, and hence make it easier to assess defects and local strains in the sample. Fig. 4c is the Fourier transform of the red-dash square region in Fig. 4b. This pattern effectively resembles the diffraction pattern of the selected region. After filtering out all other spots in the pattern, except for the red-dash circles, the reconstructed image (Fig. 4d) is the phase contrast of the vertical lattice planes. In general, the lattice planes between the silicide film and the substrate are very well aligned. Occasional mismatch at the interface also occurs, particularly around the upper corners of the pyramid regions. However, at the interface between the films and the pyramids their lattice planes are perfectly matched with each other, suggesting that the beginning of the pyramids growth is from the film. Fig. 4d also shows the strain field between the pyramid and the substrate. Lattice match between the pyramid and the substrate is found in the area closer to the film. The strain-field appears to increase further towards the pyramid's tip where the mismatch of the lattice planes is apparent as indicated by the red arrow. A similar procedure was also applied to generate the HR-TEM image featuring only the horizontal planes (Fig. 4e). The strain-field in this figure also appears to be stronger toward the pyramid's tip. The intrusion of the top of the pyramid into the film can also observed. For the top silicide film, both the vertical and horizontal lattice planes of the film are continuous and straight. Strong distortion of the horizontal planes is found mostly at the film-pyramid interface, where the pyramids intrude into the film. This observation suggests that the epitaxial film is homogeneous and single-crystalline. Our pole figure measurement of the film (data shown in Supplementary) also shows that the sample has an epitaxial cubic phase with twinning on the (111) plane. The hexagonal θ phase, which might be found in some nickel silicide films, is not recorded in this sample.

## Conclusions

In this paper, the transition process of ultrathin Ni films (∼3 nm) on Si(100) substrates are studied *in-situ* using high-resolution ion scattering (3D-MEIS) and high-resolution transmission electron microscopy. The ion scattering shows that the transformation is in three discrete steps: the as-deposited phase, the intermediate monosilicide and the final epitaxial phase. In this respect, the transition process is comparable with the conventional knowledge for the Ni-Si system. However, some significant deviations from the normal transition are found for the ultrathin films. The intermediate NiSi phase is found to occur only within a narrow range of temperature (230 − 290°C). The final epitaxial film forms at a relatively low temperature of 320 °C. Although the crystal structure of the final phase is of the *fcc* disilicide $NiSi_2$, the composition of the film is non-stoichiometric $NiSi_{1.6}$. A small fraction of the Ni films (17%) forms occasional pyramid crystal below the main silicide layer. Both ion scattering and TEM indicate the perfect alignment along the [100] axis between the film, the pyramids and the substrate. The ion scattering shows the spacing between the (100) planes of the film is found 0.7% smaller than that of the substrate.

**Acknowledgements:**

Support by VR-RFI (contracts #821-2012-5144 & #2017-00646_9) and the Swedish Foundation for Strategic Research (SSF, contract RIF14-0053 and SE13-0333) supporting accelerator operation is gratefully acknowledged. Zhen Zhang acknowledges the grant from the Swedish Foundation for Strategic Research (SSF, contract No. SE13-0333). Tuan Tran appreciates helpful discussion on the TEM data with Dr. Lars Riekehr at the Department of Electrical Engineering.

# Supplementary

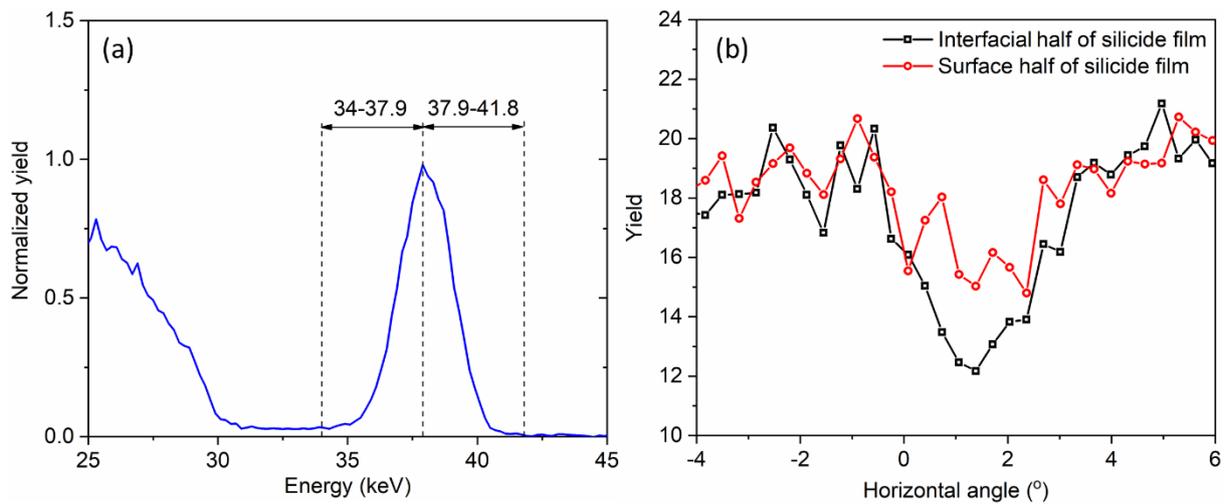

Supp. Fig. 1: Backscattering spectrum taken with a 50 keV He+ beam for the sample annealed at 290 °C (a). Sub-figure (b) is the blocking profiles within the red-dash box in Fig. 1(e) taken with the $34 - 37.9$ keV He+ ions (the interfacial half of the silicide film) and the $37.9 - 41.8$ keV He+ ions (the surface half). The stronger blocking effect shown by the black curve further indicates that the crystallization of the silicide film starts from the silicide/Si interface.

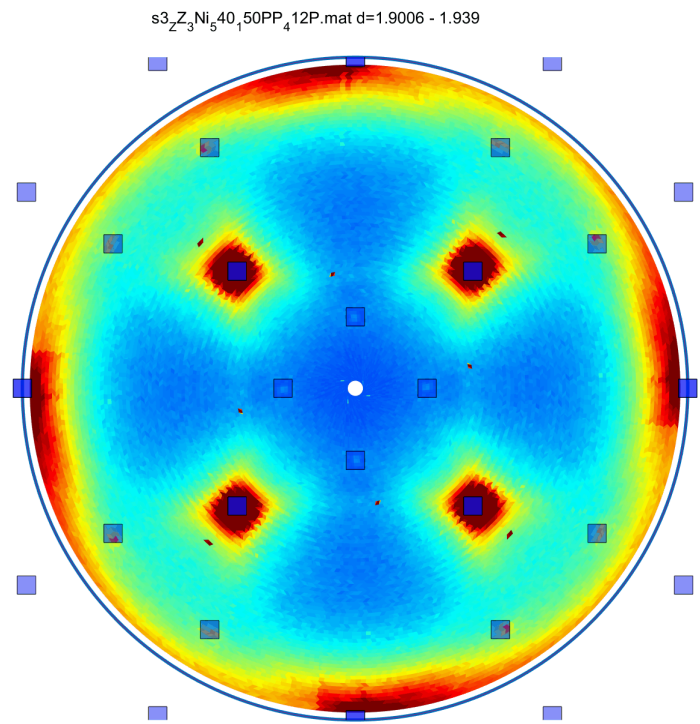

Supp. Fig. 2: Pole figure measurement on the {220} planes of the sample annealed at 540 °C. The extra dots can be fitted with a single 180 rotation on the {111} planes of a cubic structure of the size of Si or $NiSi_2$. Therefore, the pole figure measurement shows that the sample has an epitaxial cubic phase with twinning on the {111} plane. Signature spots for the hexagonal θ phase are not found in the sample by this measurement.